\documentclass[aps,twocolumn,showpacs,showkeys]{revtex4}
\usepackage{amsfonts}
\usepackage{amsmath}
\usepackage{amssymb}
\usepackage{graphicx}
\usepackage{color}
\usepackage{subfigure}
\usepackage[latin1]{inputenc}
\newcommand{\bi}{\begin{itemize}}
\newcommand{\ei}{\end{itemize}}
\newcommand{\be}{\begin{equation}}
\newcommand{\ee}{\end{equation}}
\newcommand{\ba}{\begin{eqnarray}}      
\newcommand{\ea}{\end{eqnarray}}
\newcommand{\bse}{\begin{subequations}}
\newcommand{\ese}{\end{subequations}}

\begin{document}

\title{Towards a general definition of gelation for adhesive hard-sphere dispersions}

\author{N\'estor E. Valadez-P\'erez$^{1,2}$}

\author{Yun Liu$^{2,3}$}

\author{Aaron P. R. Eberle$^{2}$}

\author{Norman J. Wagner$^{3}$}

\author{Ram\'on Casta\~neda-Priego$^{(1)}$}

\email{ramoncp@fisica.ugto.mx}

\affiliation{$^{(1)}$Divisi\'on de Ciencias e Ingenier\'ias, Campus Le\'on, Universidad de Guanajuato, Loma del Bosque 103, Lomas del Campestre, 37150 Le\'{o}n, Guanajuato, Mexico}

\affiliation{$^{(2)}$ The NIST Center for Neutron Research, National Institute of Standards and Technology, Gaithersburg, Maryland 20899-6100, USA}

\affiliation{$^{(3)}$Department of Chemical and Biomolecular Engineering, University of Delaware, Newark, Delaware 19716, USA}

\pacs{64.70.pv, 64.75.Xc, 82.70.Dd}

\keywords{Gelation, colloids, dynamical arrest}

\begin{abstract}

One major goal in condensed matter physics is identifying the physical mechanisms that lead to arrested states of matter, especially gels and glasses. The complex nature and microscopic details of each particular system are relevant. However, from both scientific and technological viewpoints, a general, consistent and unified definition is of paramount importance. Through Monte Carlo computer simulations of states identified in experiments, we demonstrate that adhesive hard-sphere dispersions are the result of rigidity percolation with average number of bonds, $\left< n_b \right>$, equals to $2.4$. This corresponds to an established mechanism leading to phase transitions in network-forming materials. Our findings connect the concept of critical gel formation in colloidal suspensions with short-range attractive interactions to the universal concept of rigidity percolation. Furthermore, the bond, angular and local distributions along the gelation line are explicitly studied in order to determine the topology of the structure of the critical gel state.

\end{abstract}

\maketitle
The enormous interest in gels and glasses resides in the fact that non-equilibrium states of matter exist in nearly every area of our daily lives and are frequently used in both technological and medical applications \citep{Zaccarelli2007}. However, due to the particular features of each system, we lack of a general definition of gelation that allows us to understand, on one hand, the route that connects a gel state with a glass transition in a continuous manner \citep{Zaccarelli2007} (and viceversa) and, on the other hand, the physical mechanisms that give rise to the formation of the arrested states of matter.

Gels and glasses typically exhibit a solid-like behavior, such as yield stress where a finite mechanical force is required to flow, but show a liquid-like (disordered) structure \citep{Zaccarelli2007,Eberle2011}. There still exists a debate with respect to what makes a gel different from a glass \citep{Zaccarelli2007,Winter2009}. One can find a large variety of properties or definitions that try to establish the differences of such non-equilibrium states. However, usually, a gel is viewed as a dilute system with a system-spanning network \citep{Zaccarelli2007,Lu2008,Laurati2009a,Eberle2011}, whereas glasses are denser systems where caging drives dynamical arrest \citep{Sciortino2002,Zaccarelli2009,Eberle2012}. These features of gels and glasses depend on the interaction potential. Therefore, the accurate knowledge of the interaction potential is needed to completely understand the mechanisms that give rise to a large diversity of arrested states. To reach this goal, one has to deal with well-controlled model systems that allow us to systematically tune the interparticle interaction \citep{Lu2008,Eberle2011,Kim2013}. 

Colloidal dispersions exhibit a rich phase behavior and undergo transitions from ergodic to non-ergodic states \citep{Solomon2010}. The most studied arrested states in systems made up of colloidal particles are the so-called repulsion-driven glasses in which colloids interact as hard-spheres \citep{Zaccarelli2009}. However, adding an attractive short-range tail to the potential enriches the landscape of the possible arrested states \citep{Zaccarelli2009}.

The inclusion of an attractive potential is essential to reach the gel transition \citep{Zaccarelli2007}. A short-range attraction between colloids can be induced, for example, by adding a non-adsorbing polymer \citep{Lu2008}; the polymer density allows us to control the attraction strength. However, to understand the effect of the induced attraction on the gel transition one has to assume that the polymer degrees of freedom can be partially mapped onto an effective interaction potential between colloids; such interaction is usually assumed of the Asakura-Oosawa form within the one-component model \citep{AO1954}. However, it has been recently discussed that to fully account for the mechanisms of dynamical arrest, one has to take into account explicitly the polymer degrees of freedom \citep{Rigo2008}.

An alternative system is the well-characterized, sterically stabilized, colloidal dispersion with adhesive interactions that are controlled via temperature to study the dynamical arrest \citep{Eberle2011,Eberle2012}. Gelation in this system is defined using the classic Winter-Chambon rheological criterion \citep{Winter1986}. A combination of small-amplitude oscillatory rheology and fiber-optic quasi-elastic light scattering is used to characterize the temperature at which dynamical arrest occurs. Small-angle neutron scattering (SANS) measurements of the structure of the dispersion at the gel temperature were carried out. The main conclusions \citep{Eberle2011,Eberle2012} are that the dynamical arrest transition in systems with short-range attractions extends from the dilute particle concentration side of the liquid-vapor coexistence to above the critical point following predictions of dynamic percolation theory, until at sufficiently high particle concentrations ($\phi\geq 0.40$, where $\phi$ is the volume fraction) it subtends the predictions and joins the mode-coupling theory (MCT) prediction for the attractive driven glass (ADG) \citep{Eberle2011,Eberle2012}.

The above explanation is based on a protocol that determines the potential parameters, i.e., attraction range and well depth, in the liquid and the arrested states. The dynamical arrest transition was characterized in terms of the Baxter parameter: $\tau^{-1}=4\left(B_2^{*}-1\right)$ \citep{Eberle2011}. This behavior is expected to be universal for all systems composed of spherical Brownian particles that interact with short-range potentials as suggested by the Noro-Frenkel (NF) extended law of corresponding states \citep{Noro2000}, which condenses all the details of the interaction potential in a single parameter, namely, the reduced second virial coefficient, $B_2^{*}(T)\equiv B_2(T)/B_2^{HS}$, where $B_2^{HS}$ is the second virial coefficient of hard-spheres. This robust description, however, did not provide a full insight of the local distribution of particles and the processes of clustering and bonding during gelation. Nevertheless, it identified the experimental boundary for gelation and, thus, allows us to explore in more detail, through theoretical tools, the mechanisms of gelation.

A common route to define a gel is through the concept of bonds \citep{Zaccarelli2007}. Particles can form bonds with a certain probability and the lifetime of bonds allows us to make a classification of gels; chemical gels are characterized by an irreversible bond formation, i.e., an infinite bond lifetime, whereas  in a physical gel bonds can reversibly break and form when the particle bonding is of the order of $k_{B}T$, i.e., the thermal energy, where $k_{B}$ is the Boltzmann's constant and $T$ the absolute temperature \citep{Zaccarelli2007}. When the attraction between particles is below $k_{B}T$, the bond lifetime is small and no physical gelation occurs, but when the attraction overcomes the thermal energy, long-lived bonds appear that can lead to the formation of a gel state. Additionally, at sufficiently low densities the latter mechanism also gives rise to a gas-liquid phase separation \citep{Zaccarelli2007,Lu2008,Eberle2011}. However, although it is clear that the formation of bonds is the main driving force to form gels, the number of bonds needed to have a stable structure capable of supporting mechanical stresses has not been established.

In this Letter we report Monte Carlo (MC) computer simulations of hard-spheres with adhesive interactions along the experimentally determined dynamical arrest transition. We demonstrate that gelation is the result of rigidity percolation that occurs when the average number of bonds takes the value of $\left< n_b \right>=2.4$. This value corresponds to that found within the context of mean-field phase transitions in random networks \citep{He1985}. Hence, this common feature opens up the possibility of introducing a unified and general definition for gelation. Further insight on the topology of the structures is gained by studying the local, bond and angular distributions along the dynamical arrest line.

Within the context of network-forming materials, covalent glasses can be divided into two classes: those with low average coordination (polymeric glasses) and those with high average coordination (amorphous) \citep{He1985,Thorpe1983}. This kind of glasses consists of rigid and floppy regions, and undergoes a mechanical phase transition as the average coordination $\left<n_{b}\right>$ (or average number of bonds) is increased and rigidity percolates through the network \citep{Thorpe1983}. He and Thorpe calculated that the phase transition takes place when the network has $\left<n_{b}\right>=2.4$ \citep{He1985}. Furthermore, a polymeric glass with both rigid and floppy regions exists for $2< \left<n_{b}\right> <2.4$. On the other hand, for $\left<n_{b}\right> >2.4$ an amorphous solid is obtained in which the rigid regions percolate. Thus, an average coordination of $2.4$ describes a rigidity percolation transition. In covalent glasses rigidity percolation leads to a permanent solid, whereas for physical bonds the solid persists on a time comparable with the bond lifetime.

The measured structure factor, $S(q)$, in the adhesive hard-sphere (AHS) dispersion was modeled by assuming a short-range square-well (SW) potential between colloids \citep{Eberle2011,Eberle2012}. According to the experimental conditions, the range, in units of the particle diameter $\sigma$, is $\lambda=1.01$ and the well depth, $\epsilon$, was adjusted to accurately describe the structure probed through SANS experiments \citep{Eberle2011,Eberle2012}. In the AHS limit, the strength of the interaction can be rewritten in terms of the $B_2^{*}(T)$ \citep{Eberle2011,Eberle2012}. As shown by Noro and Frenkel, the specific choice of potential is not important as short range attractions of less than $\sim 10\%$ follow a law of corresponding states, see, e.g., Ref. \citep{Valadez2012} and references therein.

The functional form of the SW potential unambiguously defines when two particles are linked or form a bond as when the separation between them is less or equal to the interaction range $\lambda$. Another advantage of this potential is that the total potential energy can be directly expressed in terms of the average number of bonds, i.e., $\left<U\right>=-2\left<n_{b}\right>\epsilon$. We carry out MC simulations in the $NVT$ ensemble with $N=864-4096$ particles for those states above and below the arrested states within the concentration interval: $0.11 < \phi < 0.48$. The explicit details of the simulations can be found in Ref. \citep{Valadez2012}.

\begin{figure}[t]
\centering
\includegraphics[width=0.5\textwidth]{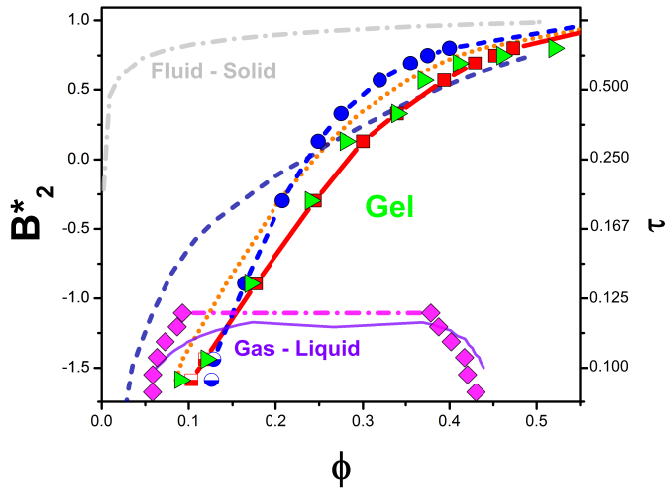} 
		\caption{State diagram for the AHS system. The solid-dotted line is the fluid-solid coexistence as determined by the self-consistent phonon theory \citep{Shin2013}. The solid line and diamonds are the gas-liquid coexistence regions for the AHS \citep{Miller2004b} and the SW fluid of range $\lambda=1.01$ (this work), respectively. The dashed line is the MCT prediction of the AHS transition \citep{Miller2004a}. The connectivity between particles is represented by the the percolation threshold (closed circles), whereas the dotted line and the line with closed squares represent those states with average coordination number values $\left<n_{b}\right>=2$ and $ \left<n_{b}\right>=2.4$, respectively. Below the gas-liquid coexistence, these states are displayed with half-filled symbols. The triangles are the experimentally determined dynamical arrest transition \citep{Eberle2011,Eberle2012}. Continuous and broken lines between symbols are to guide the eye.}
\label{fig.pd}
\end{figure}

The state diagram for the AHS dispersion is presented in Fig. \ref{fig.pd}. One can appreciate that the arrested states and the gas-liquid phase separation are buried inside the fluid-solid coexistence \citep{Shin2013}, which confirms that both phenomena occur in the meta-stable region of the diagram. Particle polidispersity frustrates crystallization in the experiments. Moreover, in the $NVT$ ensemble one is able to explore the meta-stable states in the computer simulations \citep{Valadez2012}. The gas-liquid phase coexistences predicted by Miller and Frenkel \citep{Miller2004b} and that calculated for $\lambda=1.01$ are plotted. We observe that both equilibrium diagrams are comparable. The dynamical arrest line, which corresponds to critical gel formation \citep{Winter1986}, obtained in experiments \citep{Eberle2011,Eberle2012} is also shown, along with the MCT predictions \citep{Miller2004a} and the loci of states with an average number of bonds equals to 2 and 2.4. 

At low concentrations, $\phi \lesssim 0.15$, gelation occurs inside the gas-liquid coexistence \citep{Lu2008,Eberle2011,Eberle2012,Cardinaux2007,Kim2013}. Below the binodal, percolation is congruent with gelation as observed in numerous studies \citep{Lu2008,Cardinaux2007}. The formation of a percolating network, where particles are linked by high energetic bonds ($\sim 4k_BT$), provides stability and can modify the elastic properties of the suspension. Above the critical $B_{2}^{*}$, at intermediate and high concentrations $0.15 < \phi  \lesssim 0.45$, percolation is necessary but not sufficient for gelation \citep{Coniglio2004}, as can be observed by comparing the experiments with the exact percolation line computed by MC simulations. In our previous work, it was concluded that experiments follow the percolation theory calculated from the Percus-Yevick approximation \citep{Eberle2011}. Here we find that the experiments indicate a stronger attraction is required for gelation and percolation.

At the highest concentrations $(>0.4)$, the attractive glass transition is driven by the balance between the attractive potential and the repulsion due to excluded volume effects. This balance may lead to a large variety of distinguishable nonbonded and bonded repulsive glassy states \citep{Zaccarelli2009}. Particularly, Fig. \ref{fig.pd} shows that the MCT predictions for the ADG converge to the experimental results at high concentrations $\phi>0.40$, but at lower concentrations deviations are seen. This is not unexpected as the ADG transition is a consequence of caging and this structural arrangement is only possible at higher volume fractions where there are sufficient nearest neighbors (see \citep{Zaccarelli2007} for further discussions of the limitations of the MCT theory). Therefore, strong localization of bonds must play an important role in the dynamical arrest transition below $\phi<0.40$. Moreover, $\phi_c \sim 0.4$ has been identified as the crossover from gel to glass transitions \citep{Eberle2012}. Around this point the slope of $B_2^{*}$ with $\phi$ changes significantly \citep{Eberle2012}, signalling a transition from a fluid dominated by strong bonding to a fluid dominated by excluded volume interactions augmented by weak bonding.

\begin{figure}[t]
\centering
\includegraphics[width=0.4\textwidth]{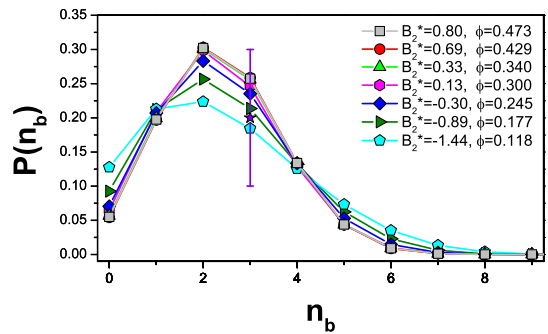}
		\caption{Bond distribution of the AHS system along the iso-coordination number curve $\left<n_{b}\right>=2.4$. The star with the error bar corresponds to the probability of finding a particle forming 3 bonds in model random networks \citep{He1985}.}
		\label{fig.coord}
\end{figure}
Of the importance in the state diagram are those states that have average bond values $2$ (dotted line) and $2.4$ (closed squares). Interestingly, the curve for $\left<n_{b}\right>=2$ closely approximates the exact percolation threshold line, whereas the curve for $\left<n_{b}\right>=2.4$ agrees quantitatively with our experimental gelation line. According to the mean-field model for network-forming materials, a system with a coordination number or average bond value of $2$ can be mechanically deformed \citep{He1985}. On the other hand, He and Thorpe demonstrated that random networks undergo a phase transition to a solid network when the $\left<n_{b}\right>=2.4$ \citep{He1985}. In order to make a straightforward comparison and analogy with the results reported in Ref. \citep{He1985}, the bond distribution for the AHS along the line $\left<n_{b}\right>=2.4$ is shown in Fig. \ref{fig.coord}. This is a non-symmetric distribution with a long tail that indicates particles are coordinated preferably with 2 to 5 particles as is typical for phase transitions of random networks \citep{He1985}. Furthermore, the distribution is nearly invariant along this iso-coordination number line despite the significant change in volume fraction. Thus, different states points $(\phi$, $B_2^{*})$ with the same average bond value would imply that the topology of the structure responsible for the critical gel is the same along the same iso-coordination number line.

\begin{figure}[t]
\centering
\includegraphics[width=0.475\textwidth]{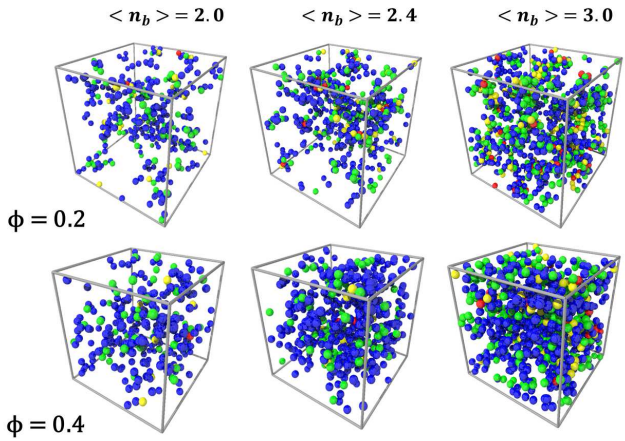}
		\caption{Snapshots of the AHS system at two different volume fractions for three different iso-coordination number curves: $\left<n_{b}\right>=2$ (left), $\left<n_{b}\right>=2.4$ (middle) and $\left<n_{b}\right>=3$ (right). Particles with 4 (blue), 5 (green), 6 (yellow) and 7 or more (red) bonds are only displayed, which would correspond to rigid regions.}
		\label{fig.snaps}
\end{figure}
He and Thorpe also calculated the probability distribution of having a coordination number of $3$ in random networks \citep{He1985}. This allows us a direct comparison with the probability of having particles forming $3$ bonds in the AHS system along the iso-coordination curve $\left<n_{b}\right>=2.4$. Remarkable, the bond distributions (see Fig. \ref{fig.coord}) lie within the values reported in Ref. \citep{He1985}. The same authors related the elastic transition in model random networks with the percolation of rigid regions, but in our simulations we cannot, strictly speaking, identify any kind of permanent rigidity because our bonds are transients. However, it is still possible to consider that particles forming 4 or more bonds are a reasonable representation of rigid regions. A visual representation of the particle distribution in real space is provided in Fig. \ref{fig.snaps}. There, snapshots at two different volume fractions for three different iso-coordination curves are presented: $\left<n_{b}\right>=2$ (left), $\left<n_{b}\right>=2.4$ (middle) and $\left<n_{b}\right>=3$ (right). The left row, almost in the percolation state, exhibits isolated regions of high coordinated or bonded particles, while middle and right rows displays a percolation of highly coordinated particles. This representation agrees with the physical picture suggested by He and Thorpe \citep{He1985} and provides insight about the formation of compact structures during gelation.

\begin{figure}[t]
\centering
\includegraphics[width=0.4\textwidth]{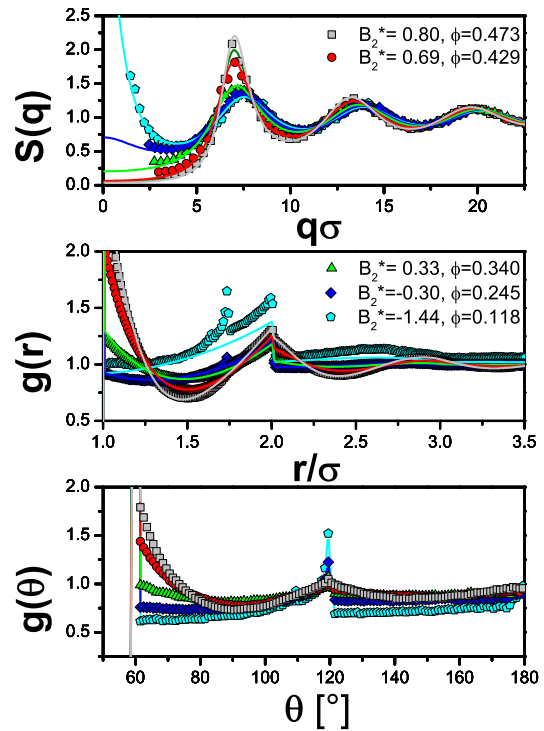}
		\caption{a) Structure factor, b) radial distribution function and c) angular bond distribution of the AHS system along the iso-bond curve $\left<n_{b}\right>=2.4$. Symbols represent simulation data and solid lines are results of the Percus-Yevick approximation \citep{Eberle2011}.}
		\label{sq.pd}
\end{figure}
To better understand the local arrangement of particles, the structure factor, the radial distribution function, $g(r)$ and the angular distribution, $g(\theta)$, along the iso-coordination curve $\left<n_{b}\right>=2.4$ are shown in Fig. \ref{sq.pd}. Here $\theta$ is defined as the relative angle formed by three bonded particles \citep{Gao2004}. The $S(q)$ exhibits a peak at $q\sigma \sim 2\pi$, which is related with the contact of particles and its height is always smaller than $2.85$ (below the Hansen-Verlet's freezing criterion \citep{Eberle2011}). However, at low-$q$ and low densities the $S(q)$ shows an upturn that is associated with the formation of large scale particle correlations; the AHS system develops long-range correlations at around and below the Boyle point ($B_2^*=0$) \citep{Lu2008,Eberle2011,Eberle2012}. At higher concentrations, such a correlation is absent as excluded volume effects dominate.

The behavior of the $g(r)$ along the gel line as shown in Fig. \ref{sq.pd} can be explained as follows. At high concentrations, corresponding to high $B_2^*$ values, the structure is similar to a hard-sphere liquid, i.e., weak bonds are present but particle correlations due to caging mechanics lead to the stability of the gel structure. On the other hand, at low volume fractions, corresponding to low $B_2^*$ values, strong bonds lead to correlations with specific angular distributions. In fact, peaks at $r/\sigma=\sqrt{3}$ are seen and can be associated with a plane trigonal particle distribution. The angular bond distribution, $g(\theta)$, shows a maximum at $60^\circ$, which indicates that particles are arranged in an equilateral triangular structure. An additional peak at $120^\circ$ is also found, which is linked to the peak found in the radial distribution at $r/\sigma=\sqrt{3}$. We do not observe the multiple peak angular distribution reported by Gao and Kilfoi for colloid-polymer mixtures \citep{Gao2004}. This difference might be due to the more complex nature of a two-component system.

In conclusion, by means of Monte Carlo computer simulations and experiments of nanoparticle dispersions with short-range interactions, we have demonstrated that critical gel formation in AHS dispersions is the result of rigidity percolation of a dynamic network with an average value $\left< n_b \right>=2.4$. Thus, dynamical arrest for AHS dispersions is another example of a rigidity phase transition in network-forming materials. Furthermore, our findings open up the possibility of having a general definition of gelation in attractive-driven colloidal suspensions. From the scientific point of view, this discovery establishes a consistent and unified definition of critical gel formation. Finally, theses findings can facilitate the quantitative prediction of important product properties for the manufacturing, fabrication and processing of commercial products based on gels.

\begin{acknowledgments}
R.C.P. thanks to Adri\'an Huerta for useful discussions on rigidity percolation. This work was financial supported by CONACyT (grant 102339/2008) and NSF-CONACyT (project 147892/2011). The funding for A.P.R.E. was provided by the National Academy of Science through a National Research Council postdoctoral fellowship. This manuscript was prepared under cooperative agreement 70NANB7H6178 from NIST, U.S. Department of Commerce. The statements, findings, conclusions and recommendations are those of the author(s) and do not necessarily reflect the view of NIST or the U.S. Department of Commerce.
\end{acknowledgments}

\end{document}